\documentclass[a4paper,10pt]{article}
\usepackage{algorithmic,algorithm}
\usepackage{color,colortbl,float,subfig,multicol,textcomp}
\usepackage[colorlinks,citecolor=blue,linkcolor=black]{hyperref}
\usepackage{amsmath,amsfonts,amssymb,graphicx,latexsym}
\usepackage[a4paper,includeheadfoot,pdftex]{geometry}%, top = 2cm, bottom = 2cm,
\newcommand{\bra}[1]{\left\langle{#1}\right\vert} %Dirac ket
\newcommand{\ket}[1]{\left\vert{#1}\right\rangle} % Dirac bra
\newcommand{\opn}[1]{\ensuremath{\operatorname{#1}}} % Operatorname
\title{Quantum associative memory for the diagnosis of some tropical diseases}
\author{J-P. TCHAPET NJAFA, S.G. NANA ENGO \\
\small{Laboratory of Photonics, Department of Physics,} \\\small{University of
Ngaoundere, POB 454 Ngaoundere, Cameroon}\\
              \\
           P. WOAFO \\
\small{              Laboratory of Modelling and Simulation in Engineering and
Biological Physics,}\\\small{ Department of Physics, Faculty of Science,}\\
\small{University of Yaounde I, POB 812 Yaounde, Cameroon}
}

\begin{document}
\maketitle
\begin{abstract}
In this paper we present a model of Quantum Associative Memory which can be a
helpful tool for physicians without experience or laboratory facilities, for the
diagnosis of four tropical diseases (malaria, typhoid fever, yellow fever and
dengue) which have similar symptoms. The memory can distinguish single infection
from multi-infection. The algorithm used for Quantum Associative Memory is an
improve model of original algorithm made by Ventura for Quantum Associative
Memory. From the simulation results given, it appears that the efficiency of
recognition is good when a particular symptom of a disease with a similar 
symptoms
are inserted.
\end{abstract}

\section{Introduction}
Diagnosis is the identification of abnormal condition that afflicts a specific 
patient, based on manifested clinical data or lesions. If the final diagnosis 
agrees with a disease that afflicts a patient, the diagnostic process is 
correct; otherwise, a misdiagnosis occurred. Medical diagnosis is a 
categorization task that allows physicians to make prediction about features of 
clinical situations and to determine appropriate course of action. It involves a 
complex decision process that involves a lot of vagueness and uncertainty 
management, especially when the disease has multiple symptoms.

Artificial neural networks provide a powerful tool to help physicians avoiding 
misdiagnosis by analyzing, modelling and making sense of complex clinical data 
across a broad range of medical applications. Most of medical applications of 
artificial neural networks are classification problems; that is, the task is on 
the basis of the measured features to assign the patient to one of a small set 
of classes \cite{Dybowski}.

Malaria is most world parasitic disease. $40\%$ of the world's population are 
concerned, especially those of tropical regions. In Cameroon it is a public 
health problem because all the population is exposed to the disease. To diagnose 
malaria, the World Health Organization (WHO) \cite{who} recommends the use of 
rapid diagnostic testing. But these tools need some conservations facilities 
which are difficult to find in rural and semi-urban regions in developing 
countries. In fact it is crucial to maintain the frozen chain because the 
storage over some temperatures and high humidity affect their sensibility and 
efficiency. So the most widely used technique for determining the development 
stage of the malaria disease is visual microscopical evaluation of Giemsa 
stained blood smears. However, this is a routine and time-consuming task and 
requires a trained operator. But, most of the time there is a misdiagnosis 
because of confusion between symptoms of malaria and symptoms of other tropical 
diseases like typhoid fever, yellow fever and dengue, or the inexperience of the 
physicians.

For the diagnosis of four tropical diseases (malaria, typhoid fever, yellow 
fever and dengue) which have similar symptoms, the purpose the Quantum 
Associative Memory (QAM) proposed here is to (i) act as an advisory tool to 
novice users, specifically senior nurses in rural health centers with limited or 
no physicians; (ii) act as a decision support tool for medical diagnosis for 
physicians in under staffed health centers; (iii) provide an alternative way to 
reach a reasonable tentative diagnosis, and hence early commencement of clinical 
management of patients in the absence of laboratory facilities in many rural and 
semi-urban health centers.

Some computer-assisted tools in the case of tropical diseases are already built 
and some of them use artificial neural networks, but they are only specialized 
for malaria  \cite{Sunny,neetu,Khalda,andrade,Filippo}. The use of artificial 
neural networks is motivated by the fact that they can capture domain knowledge 
from example and have good generalization. Our model, which generalize the work 
of Agarkar and Ghathol \cite{Agarkar} that use the FFANN for the diagnosis of 
malaria, typhoid fever and dengue, will be further extended to a wide range of 
tropical diseases. 

The paper is structured as follows. Section \ref{sec:dis} provides a brief 
description of the symptoms of each disease, described in detail in Appendix 
\ref{sec:signs}. In Section \ref{sec:meth}, foundations of our Quantum 
Associative Memory are presented. Section \ref{sec:Sim} is devoted to 
simulations and results. Finally, we conclude with an outlook of possible 
future directions.

\section{Short description of the symptoms of the diseases \cite{epilly,dico}} 
\label{sec:dis}

We briefly give here the symptoms of each disease, described in detail in 
Appendix \ref{sec:signs}.

\begin{description}
\item[Malaria] is caused by protozoan parasites of the genus \emph{plasmodium}. 
Four species infect humans by entering the bloodstream: \emph{P. vivax, P. 
ovale, P. malariae and P. falciparum} which affects a greater proportion of the 
red blood cells than the others and is most serious. The parasite is generally 
transmitted from one human to another by the bite of infected anopheles 
mosquitoes.

\item[Typhoid fever,] also known as salmonellosis, is a bacterial disease caused 
by \emph{sallmonella typhi}. Dirt is the main cause of transmission of the 
disease. Contaminated food and water are the principal vectors.

\item[Yellow fever] is caused by a \emph{flavivirus}. The virus is transmitted 
by the bite of a mosquito (\emph{Aedes aegypti}). The virus cause deterioration 
of the liver.

\item[Dengue] is caused by \emph{dengue virus}. The virus is transmitted by the 
bite of mosquito (\emph{Aedes aegypti} and \emph{Aedes albopictus}). The WHO 
has classified dengue as one of the neglected tropical diseases and report the 
resurgence of the disease \cite{neglige}.
\end{description}

\section{Method}\label{sec:meth}

\subsection{Algorithm for data search}
The algorithm that we use for retrieving was originally proposed by Ezhov and 
Ventura \cite{ezhov2000} as \emph{quantum associative memory with distributed 
queries}. In the Ref. \cite{tchapet2012} we improved it by making some 
modifications. The resulting algorithm is given by Algorithm \ref{alg:requetm}.

\begin{algorithm}[H]
\caption{Improve Quantum Associative Memory with distributed query}
\label{alg:requetm}
\begin{algorithmic}[1]
\STATE $\ket{0_10_2\dots 0_n}\equiv \ket{\bar{0}}$;
\COMMENT{Initialize the register}
\STATE $\ket{\Psi}=A\ket{\bar{0}}=\frac{1}{\sqrt{N-m}}\sum^{N-1}_{x\notin M} 
\ket{x}$;
\COMMENT{Learn the patterns using exclusion approach which can be made by the
Binary Superposed Quantum Decision Diagram (BSQDD) proposed by Rosenbaum (see
\cite{Rosenbaum2010} for detail)}
\STATE \label{etapp3} Apply the operator oracle $\mathcal{O}$ to the register;
\STATE Apply the operator diffusion $\mathcal{D}$ to the register;
\STATE \label{etapp5} Apply operator $\mathcal{I}_M$ to the register;
\STATE \label{etapp6} Apply the operator diffusion $\mathcal{D}$ to the 
register;
\REPEAT
\STATE \label{etap6} Apply the operator oracle $\mathcal{O}$ to the register;
\STATE \label{etap7} Apply the operator diffusion $\mathcal{D}$ to the register;
\STATE $i=i+1$;
\UNTIL{$i>\Lambda-2$}
\STATE Observe the system.
\end{algorithmic}
\end{algorithm}

It should be noted that to get original Ezhov and Ventura's algorithm 
\cite{ezhov2000} lines $3$ sto $6$ must be omitted and in line $11$ $i>\Lambda$.

The model uses the \emph{exclusion learning approach} in which the system is in 
the superposition of all the possible states, except the patterns states. If 
$M$ is the set of patterns and $m$ the number of patterns of length $n$,
\begin{equation}
 \ket{\Psi}=\frac{1}{\sqrt{N-m}}\sum^{N-1}_{x\notin M}\ket{x},\,N=2^n.
\label{eq:Psi}
\end{equation}
In other words, the exclusion approach for the learning pattern included each
point not in $M$ with nonzero coefficient while those points in $M$ have zero
coefficients.

The distributed query is in the following superposed states
\begin{equation}
 \ket{Req^p}=\sum^{N-1}_{x=0}Req^p_x\ket{x},
\end{equation}
where $Req^p_x$ obey to binomial distribution
\begin{equation}
 \|Req^p_x\|^{2}=a^{d_H(p,x)}(1-a)^{n-d_H(p,x)}.
\label{eq:re}
\end{equation}
In equation (\ref{eq:re})
\begin{itemize}
 \item $p$ marks the state $\ket{p}$ which is referred as the query center;
\item $0<a<\frac{1}{2}$ is an arbitrary value that regulates the width of the
distribution;
\item the \textbf{Hamming distance} $d_H(p,x)=|p-x|$ between $\ket{x}$ and
$\ket{p}$ is an important tool which gives the correlation between input and
output;
\item the amplitudes are such that $\sum_x\|Req^p_x\|^{2}=1$.
\end{itemize}

In the Algorithm \ref{alg:requetm},
\begin{itemize}
 \item $\mathcal{O}$ is the operator oracle which inverts the phase of the
query state $\ket{Req^p}$,
 \begin{align}
 \mathcal{O}& =\mathbb{I}-(1-e^{i\pi})\ket{Req^p}\bra{Req^p},\\
  \mathcal{O} &:a_x\mapsto a_x-2Req^p_x\left(\sum^{2^{n}-1}_{x=0}
(Req^p_x)^{*}a_x\right),
\label{eq:O}
\end{align}
where $a_x$ is the probability amplitude of the state $\ket{x}$.

\item $\mathcal{D}$ is the operator diffusion which inverts the probability
amplitude of the states of $\ket{\Psi}$ over their average amplitude and for the
others over the value $0$.
 \begin{align}
  \mathcal{D}& =(1-e^{i\pi})\ket{\Psi}\bra{\Psi}-\mathbb{I},\\
 \mathcal{D} &:a_x\mapsto 2m_x\left(\sum^{N-1}_{x=0}m^{*}_xa_x\right)-a_x.
\label{eq:D}
\end{align}
where $m_x$ is the probability amplitude of a state of
$\ket{\Psi}$.
\item $\Lambda$ is the number of iterations that yields the maximal value of
amplitudes, which must be as far as possible nearest to an integer,
\begin{equation}
 \Lambda=T(\frac{1}{4}+\alpha),\,T=\frac{2\pi}{\omega},\,\alpha\in\mathbb{N},
\label{eq:lambda}
\end{equation}
with the Grover's frequency
\begin{equation}
\label{equaB}
 \omega=2\arcsin{B},\,B=\frac{1}{\sqrt{N-m}}\sum^{N-1}_{x=0,x\notin M}Req^p_x.
\end{equation}
\end{itemize}
Two cases we be will considered for the operator $\mathcal{I}_M$:
\begin{itemize}
\item[\textbf{QAM-C1}:] $\mathcal{I}_M$ inverts only the phase of the memory
patterns states as in the Ventura's model,
\begin{align}
 \mathcal{I}_M &=\mathbb{I}-(1-e^{i\pi})\ket{\varphi}\bra{\varphi},\,
\ket{\varphi}\bra{\varphi}=\sum_{x\in M}\ket{x}\bra{x},\\
 \mathcal{I}_M & :a_x\mapsto
\begin{cases}
-a_x\text{ if }\ket{x}\in M\\
a_x\text{ if not.}
\end{cases}
\end{align}
$\forall x\in M$, the Grover operator acts as
\begin{align}
 \begin{split}
\mathcal{D}\mathcal{I}_M\ket{\varphi}&=(2\ket{\Psi}\bra{\Psi}-\mathbb{I}
+2\ket{\varphi}\bra{\varphi})\ket{\varphi}\\
& =2\ket{\Psi}\langle\Psi\ket{\varphi}-\ket{\varphi}+2\ket{\varphi}
\langle\varphi\ket{\varphi}=\ket{\varphi}.
 \end{split}
\end{align}

 \item[\textbf{QAM-C2}:] $\mathcal{I}_M$ is formally identical to the operator
oracle $\mathcal{O}$ of Eq. (\ref{eq:O}),
\begin{equation}
  \mathcal{I}_M :a_x\mapsto a_x-2REQ_x\left(\sum^{N-1}_{x=0}
(REQ_x)^{*}a_x\right),
\end{equation}
with
\begin{equation}
 \label{eq:RE}
\|REQ_x\|^{2} =\frac{1}{k}\sum_{p}a_{b}^{d_H(b,x)}(1-a_{b})^{n-d_H(b,x)},
\end{equation}
where we consider that the distribution has $k$ centers and $0<a_b<\frac{1}{2}$ 
is an arbitrary value that regulates the width distribution around the center 
$b$. But in this paper, we will consider that
\begin{equation}
 \label{eq:REm}
\|REQ_x\|^{2}=\frac{1}{m}\sum_{b \in M}a'^{d_H(b,x)}(1-a')^{ n-d_H(b,x)}.
\end{equation}
where $m$ is the number of patterns for the learning, $b$ is an item of the set
$M$ of patterns, and we choose the case where $a'=a_{b}$ is the same for all the
patterns.
\end{itemize}

\subsection{Database}
Our database contains symptoms of four tropical diseases: malaria, typhoid 
fever, dengue and yellow fever. There is $23$ symptoms for malaria, $16$ for 
typhoid fever and dengue and $10$ for yellow fever. According to the fact that 
some symptoms are common to different diseases, not only the four, like fever 
and headache for example, our database is reduced to $47$ symptoms. Each symptom 
is labelled with a number, from $0$ to $46$, in his binary form, so we need $6$ 
qubits for the computation. Furthermore, we need $4$ qubits to label the ten 
groups of diseases presented by Table \ref{tab:indexDisease}. There is one group 
per individual disease (N\textdegree{} 1-4); three groups corresponding to 
common symptoms to malaria and typhoid fever, or dengue, or yellow fever 
(N\textdegree{} 5-7); one group corresponding to common symptoms to yellow fever 
and dengue (N\textdegree{} 8); one group corresponding to common symptoms to 
malaria, yellow fever and dengue (N\textdegree{} 9); and finally one group 
corresponding to common symptoms to each of the four diseases and that can be 
found in other diseases as headache, fever and vomiting (N\textdegree 10). It 
appears that we need a register which contains $n=10$ qubits, $6$ for symptoms 
and $4$ for diseases, all labelled as mentioned above. The other possibilities 
are pointed to be other diseases and symptoms in our model. 

\begin{table}[hbtp]
\centering
\begin{tabular}{llc}\hline
\textbf{N\textdegree} &\textbf{Group of diseases by symptoms} & 
\textbf{Label}\\\hline
1 & Malaria & $\ket{0001}$\\%\hline
2 & Typhoid fever & $\ket{0010}$\\%\hline
3 & Yellow fever & $\ket{0100}$\\%\hline
4 & Dengue & $\ket{1000}$\\%\hline
5 & Malaria + Typhoid fever & $\ket{0011}$\\%\hline
6 & Malaria + Dengue & $\ket{1001}$\\%\hline
7 & Malaria + Yellow fever & $\ket{0101}$\\%\hline
8 & Yellow fever + Dengue & $\ket{1100}$\\%\hline
9 & Malaria + Yellow fever + Dengue & $\ket{1101}$\\%\hline
10 & Other diseases & $\ket{0000}$\\\hline
\end{tabular}
\caption{Groups of diseases by symptoms and their labels in binary form. The 
hamming distance between the label of two groups of diseases is equal to one 
when the symptoms are common to these two groups of diseases and is equal to 
two otherwise.The group N\textdegree{} 10 or \emph{Other groups of diseases} is 
devoted to symptoms that are common to each of the four diseases and that can 
also occur in other groups of diseases which are not mentioned here.}
\label{tab:indexDisease}
\end{table}

As we want the Quantum Associative Memory to give as output a disease that 
corresponds to symptoms give as inputs, some modifications must be done on the 
operators of the Algorithm \ref{alg:requetm}. Concretely, we want the operator 
$\mathcal{O}$ and the operator $\mathcal{I}_M$ to act on the subspace of 
diseases (last $4$ qubits): $\mathbb{I}_6\otimes\mathcal{O}_4=O$; while the 
operator $\mathcal{D}$ acts on the subspace of symptoms (first $6$ qubits): 
$\mathcal{D}_6\otimes\mathbb{I}_4=D$. The Algorithm \ref{alg:requetm} must be 
rewriting as the Algorithm \ref{alg:requetm2}, where $\mathbb{I}$ is the 
identity operator. The determination of the number of iterations is still the 
same as mentioned above, but the distributed query must be transposed from his 
subspace to the global Hilbert's space as
\begin{equation}
\ket{Req^p}=\frac{1}{8}\sum^{15}_{y=0}\sum^{63}_{x=0}Req^p_y\ket{x}\otimes\ket{y
}, \label{eq:req}
\end{equation}
where $p$ takes one of decimal values of the label gives on the Table 
\ref{tab:indexDisease}.

\begin{algorithm}[htpb]
\caption{Improve Quantum Associative Memory with distributed query for
diagnosis \#1}
\label{alg:requetm2}
\begin{algorithmic}[1]
\STATE $\ket{0_10_2\dots 0_n}\equiv \ket{\bar{0}}$;
\COMMENT{Initialize the register}
\STATE $\ket{\Psi}=A\ket{\bar{0}}=\frac{1}{\sqrt{N-m}}\sum^{N-1}_{x\notin
M}\ket{x}$;
\COMMENT{Learn the patterns using exclusion approach which can be made by the
Binary Superposed Quantum Decision Diagram (BSQDD) proposed by Rosenbaum (see
\cite{Rosenbaum2010} for detail)}
\STATE \label{etapp32} Apply the operator oracle
$\mathbb{I}_6\otimes\mathcal{O}_4=O$ to the register;
\STATE Apply the operator diffusion $\mathcal{D}_6\otimes\mathbb{I}_4=D$ to the
register;
\STATE \label{etapp52} Apply operator $\mathbb{I}_6\otimes\mathcal{I}_{M4}=I_M$
to the register;
\STATE \label{etapp62} Apply the operator diffusion 
$\mathcal{D}_6\otimes\mathbb{I}_4=D$ to the register;
\REPEAT
\STATE \label{etap62} Apply the operator oracle 
$\mathbb{I}_6\otimes\mathcal{O}_4=O$ to the register;
\STATE \label{etap72} Apply the operator diffusion
$\mathcal{D}_6\otimes\mathbb{I}_4=D$ to the register;
\STATE $i=i+1$;
\UNTIL{$i>\Lambda-2$}
\STATE Observe the system.
\end{algorithmic}
\end{algorithm}

\section{Simulations and results}\label{sec:Sim}

For the simulations we choose four symptoms that we know the corresponding 
disease, i.e., symptoms that occur only when a patient suffers from the 
corresponding disease. As for \textbf{QAM-C2} method, best results appear when 
$a'$ is close to $0.5$, we take here $0.4999$. The symptoms and an appropriate 
scheme determine the center of the query. For example, if we introduce some 
symptoms of malaria and typhoid fever, the center of the query will be 
$\ket{0011}$. We give the ratio $P_c/P_w$ which is the recognition efficiency of 
the Quantum Associative Memories, where $P_c $ and $P_w$ are probabilities of 
correct and incorrect recognition. More is this ratio, better is the 
recognition.
\begin{enumerate}

 \item \textbf{Malaria}: Table \ref{tab:resultMala} and Figure 
\ref{fig:resultMala} give the relevant parameters of each method.
 
 \begin{table}[hbtp]
  \begin{center}
  {%
\newcommand{\mc}[3]{\multicolumn{#1}{#2}{#3}}
\begin{center}
\begin{tabular}{llcc}\hline
QAM & $\Lambda$ & \mc{2}{c}{$P_c/P_w$}\\\cline{3-4}
 &  & Measure on the $6$ first qubits & Measure on the $4$ last qubits\\\hline
Ezhov's & $1$ & $0.1767$ & $1.5823$\\
\textbf{C1} & $2$ & $0.1873$ & $1.9710$\\
\textbf{C2} & $2$ & $0.1841$ & $1.6900$\\\hline
  \end{tabular}
  \end{center}
}%
\caption{Relevant parameters of Ezhov's, \textbf{QAM-C1} and \textbf{QAM-C2} 
methods in case of the diagnosis of malaria.}
\label{tab:resultMala}
\end{center}
 \end{table}
 
\begin{figure}[hbtp]
\begin{center}
\leavevmode
 \subfloat[Ezhov's]{
 \includegraphics[scale=0.8]{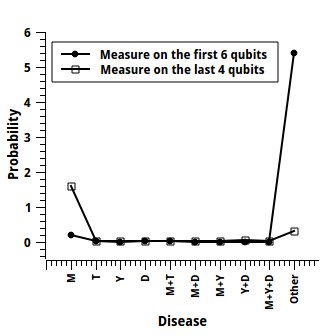}}%\vspace{0.5cm}
 \subfloat[\textbf{QAM-C1}]{
 \includegraphics[scale=0.8]{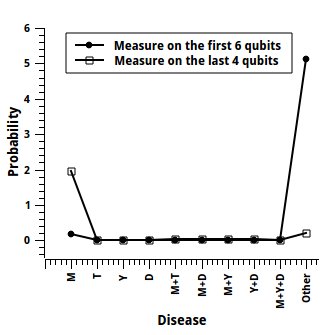}}\\%\hspace{0.5cm}
 \subfloat[\textbf{QAM-C2}]{
 \includegraphics[scale=0.8]{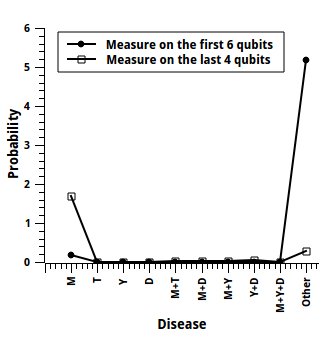}}
 \caption{Recognition efficiency of the diagnosis in case of malaria.
M: malaria; T: typhoid fever; Y: yellow fever; D: dengue; Other: other disease. 
The line with circles is the recognition efficiency when the measure is done
on the first $6$ qubits (symptoms) while the line with empty square is the
recognition efficiency when the measure is done on the last $4$ qubits
(disease)).}
\label{fig:resultMala}
\end{center}
\end{figure}

\item \textbf{Typhoid fever or dengue}: Table \ref{tab:resultTypho} and
Figure \ref{fig:resultTypho} give the relevant parameters of each method for 
the diagnosis of typhoid fever. In case of the diagnosis of dengue, same 
results 
occur. The reason is the number of symptoms of each disease which is equal for 
both.
 \begin{table}[hbtp]
  \begin{center}
  {%
\newcommand{\mc}[3]{\multicolumn{#1}{#2}{#3}}
\begin{center}
\begin{tabular}{llcc}\hline
QAM & $\Lambda$ & \mc{2}{c}{$P_c/P_w$}\\\cline{3-4}
 &  & Measure on the $6$ first qubits & Measure on the $4$ last qubits\\\hline
Ezhov's & $11$ & $0.1374$ & $1.2967$\\
\textbf{C1} & $14$ & $0.1374$ & $1.2967$\\
\textbf{C2} & $12$ & $0.1611$ & $1.4125$\\\hline
  \end{tabular}
  \end{center}
}%
\caption{Relevant parameters of Ezhov's, \textbf{QAM-C1} and \textbf{QAM-C2} 
methods in case of the diagnosis of typhoid fever (or dengue).}
\label{tab:resultTypho}
\end{center}
 \end{table}
 
\begin{figure}[hbtp]
\begin{center}
\leavevmode
 \subfloat[Ezhov's and \textbf{QAM-C1}]{
 \includegraphics[scale=0.8]{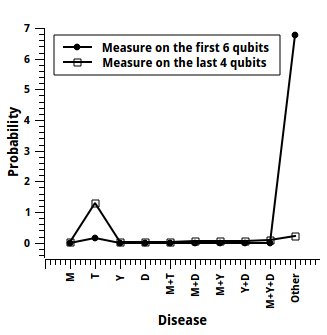}}\vspace{0.5cm}
 \subfloat[\textbf{QAM-C2}]{
 \includegraphics[scale=0.8]{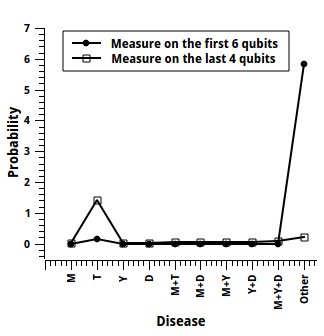}}
 \caption{Recognition efficiency of the diagnosis in case of typhoid fever or
dengue. M: malaria; T: typhoid fever; Y: yellow fever; D: dengue; Other: other 
disease. The line with circles is the recognition efficiency when the measure 
is 
done on the first $6$ qubits (symptoms) while the line with empty square is the 
recognition efficiency when the measure is done on the last $4$ qubits 
(disease)).}
\label{fig:resultTypho}
\end{center}
\end{figure}

\item \textbf{Yellow fever}: in this case we choose only two symptoms of the
disease because it has the lower number of symptoms. Here Table
\ref{tab:resultYello} and Figure \ref{fig:resultYello} give the relevant 
parameters of each method.
\begin{table}[hbtp]
  \begin{center}
  {%
\newcommand{\mc}[3]{\multicolumn{#1}{#2}{#3}}
\begin{center}
\begin{tabular}{llrl}\hline
QAM & $\Lambda$ & \mc{2}{c}{$P_c/P_w$}\\\cline{3-4}
 &  & Measure on the $6$ first qubits & Measure on the $4$ last qubits\\\hline
Ezhov's & $16$ & $0.0638$ & $0.7780$\\
\textbf{C1} & $19$ & $0.0638$ & $0.7780$\\
\textbf{C2} & $13$ & $0.0880$ & $0.9550$\\\hline
  \end{tabular}
  \end{center}
}%
\caption{Relevant parameters of Ezhov's, \textbf{QAM-C1} and \textbf{QAM-C2} 
methods in case of the diagnosis of yellow fever.}
\label{tab:resultYello}
\end{center}
 \end{table}
 
\begin{figure}[hbtp]
\begin{center}
\leavevmode
 \subfloat[Ezhov's and \textbf{QAM-C1}]{
 \includegraphics[scale=0.8]{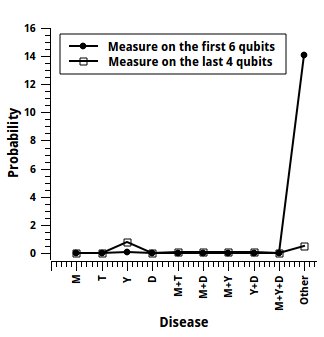}}\vspace{0.5cm}
 \subfloat[\textbf{QAM-C2}]{
 \includegraphics[scale=0.8]{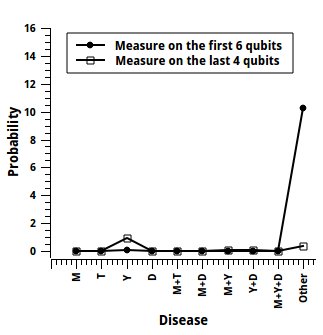}}
 \caption{Recognition efficiency of the diagnosis in case of yellow fever. M: 
malaria; T: typhoid fever; Y: yellow fever; D: dengue; Other: other disease. 
The 
line with circles is the recognition efficiency when the measure is done on the 
first $6$ qubits (symptoms) while the line with empty square is the recognition 
efficiency when the measure is done on the last $4$ qubits (disease)).}
\label{fig:resultYello}
\end{center}
\end{figure}
\end{enumerate}

As we see on each figure, due by the hamming distance, the Quantum Associative 
Memory can collapse to the states of mark for multi-infection. But the use of 
the \textbf{QAM-C2} method seem to be the best in terms of probabilities and 
iterations for single disease retrieving. The Quantum Associative Memory is 
also 
built to identify a disease with the lowest (one) or the highest (all) number 
of 
particular symptoms of this disease. But when we introduce one or more of the 
three common symptoms, the probability that the Quantum Associative Memory 
collapses to other disease is still important. Table \ref{tab:probaMa} shows it 
when the same symptoms previously introduce are completed with the common 
symptoms in case of malaria when measure was done on last four qubits.
\begin{table}[hbtp]
\begin{center}
{%
\newcommand{\mc}[3]{\multicolumn{#1}{#2}{#3}}
\begin{center}
\begin{tabular}{cccc}\hline
Number of & \mc{2}{c}{$P_c/P_w$} & $\Lambda$\\\cline{2-3}
common symptoms & \mc{1}{r|}{Malaria} & \mc{1}{l}{Other diseases} & \\\hline
$0$ & \mc{1}{r|}{$1.6900$} & \mc{1}{l}{$0.2775$} & $2$\\
$1$ & \mc{1}{r|}{$2.0736$} & \mc{1}{l}{$0.1391$} & $11$\\
$2$ & \mc{1}{r|}{$0.9180$} & \mc{1}{l}{$0.4712$} & $12$\\
$3$ & \mc{1}{r|}{$1.3156$} & \mc{1}{l}{$0.7503$} & $9$\\\hline
\end{tabular}
\end{center}
}%
\caption{Recognition efficiency in case of malaria when the common symptoms are
introduce.}
\label{tab:probaMa}
\end{center}
\end{table}

As we build the Quantum Associative Memory it collapses, with a good 
probability, to a correct state of a particular disease when only a few 
symptoms 
or all symptoms of this disease are introduced. But, generally it is the common 
symptoms which are easily identified with some particular symptoms and it is 
not 
easy to find material to identify the other particular symptoms in rural or 
semi-urban health center. It will be great to have a Quantum Associative Memory 
which can identify an infection or multi-infection with the lowest rate of 
particular symptoms and the common symptoms.

One way to get this issue is to invert the amplitudes of probability of the 
states of symptoms which are not excluded during the learning step, after the 
identification of the center of query. Remember that the center of the query is 
one of the state gave in Table \ref{tab:indexDisease} which represents the 
diseases. For this issue, the Algorithm \ref{alg:requetm2} will be rewriting as 
Algorithm \ref{alg:requetm3}.
\begin{algorithm}[htpb]
\caption{Improve Quantum Associative Memory with distributed query for
diagnosis \#2}
\label{alg:requetm3}
\begin{algorithmic}[1]
\STATE $\ket{0_10_2\dots 0_n}\equiv \ket{\bar{0}}$;
\COMMENT{Initialize the register}
\STATE $\ket{\Psi}=A\ket{\bar{0}}=\frac{1}{\sqrt{N-m}}\sum^{N-1}_{x\notin
M}\ket{x}$;
\COMMENT{Learn the patterns using exclusion approach which can be made by the
Binary Superposed Quantum Decision Diagram (BSQDD) proposed by Rosenbaum (see
\cite{Rosenbaum2010} for detail)}
\STATE \COMMENT{Invert the amplitudes of probability of states corresponding to
symptoms of the center which are not excluded.}
\STATE \label{etapp33} Apply the operator oracle
$\mathbb{I}_6\otimes\mathcal{O}_4=O$ to the register;
\STATE Apply the operator diffusion $\mathcal{D}_6\otimes\mathbb{I}_4=D$ to the
register;
\STATE \label{etapp53} Apply operator
$\mathbb{I}_6\otimes\mathcal{I}_{M4}=I_M$
to the register;
\STATE \label{etapp63} Apply the operator diffusion
$\mathcal{D}_6\otimes\mathbb{I}_4=D$ to the register;
\REPEAT
\STATE \label{etap63} Apply the operator oracle
$\mathbb{I}_6\otimes\mathcal{O}_4=O$ to the register;
\STATE \label{etap73} Apply the operator diffusion
$\mathcal{D}_6\otimes\mathbb{I}_4=D$ to the register;
\STATE $i=i+1$;
\UNTIL{$i>\Lambda-2$}
\STATE Observe the system.
\end{algorithmic}
\end{algorithm}

With the same data used to build the Table \ref{tab:probaMa} with the 
\textbf{QAM-C2} method, we obtain the Table \ref{tab:probaMa1}.

\begin{table}[htpb]
\begin{center}
{%
\newcommand{\mc}[3]{\multicolumn{#1}{#2}{#3}}
\begin{center}
\begin{tabular}{cccc}\hline
Number of & \mc{2}{c}{$P_c/P_w$} & $\Lambda$\\\cline{2-3}
common symptoms & \mc{1}{r|}{Malaria} & \mc{1}{l}{Other diseases} & \\\hline
$0$ & \mc{1}{r|}{$2.4397$} & \mc{1}{l}{$0.2067$} & $15$\\
$1$ & \mc{1}{r|}{$2.2376$} & \mc{1}{l}{$0.1391$} & $2$\\
$2$ & \mc{1}{r|}{$1.5984$} & \mc{1}{l}{$0.3925$} & $16$\\
$3$ & \mc{1}{r|}{$10.8297$} & \mc{1}{l}{$0.0794$} & $33$\\\hline
\end{tabular}
\end{center}
}%
\caption{Recognition efficiency in case of malaria when the common symptoms are
introduce and the phase inversion.}
\label{tab:probaMa1}
\end{center}
\end{table}
The fewest of particular symptoms and the great majority of common symptoms 
(especially fever, headache and vomiting) give the best probability of 
retrieving. Let us see it with two particular cases already mentioned with the 
\textbf{QAM-C2} method.
\begin{enumerate}
 \item Only one particular symptom of malaria is introduced and the three common
symptoms (Table \ref{tab:resul} and Figure \ref{fig:resul}).
 \begin{table}[htpb]
  \begin{center}
  {%
\newcommand{\mc}[3]{\multicolumn{#1}{#2}{#3}}
\begin{center}
\begin{tabular}{llrl}\hline
Method & $\Lambda$ & \mc{2}{c}{$P_c/P_w$}\\\cline{3-4}
 &  & Measure on the $6$ first qubits & Measure on the $4$ last qubits\\\hline
with phase inversion & $10$ & $6.5210$ & $15.1820$\\
without phase inversion & $13$ & $0.3321$ & $0.3325$\\\hline
  \end{tabular}
  \end{center}
}%
\caption{Relevant parameters of the \textbf{QAM-C2} method in case of the 
diagnosis of malaria.}
\label{tab:resul}
\end{center}
 \end{table}
\begin{figure}[htp]
\begin{center}
\leavevmode
 \subfloat[with phase inversion]{
 \includegraphics[scale=0.8]{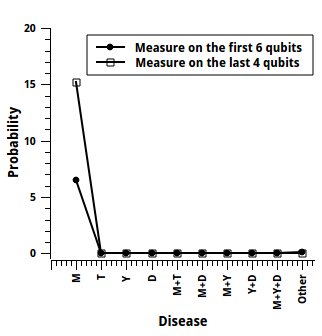}}\vspace{0.5cm}
 \subfloat[without phase inversion]{
 \includegraphics[scale=0.8]{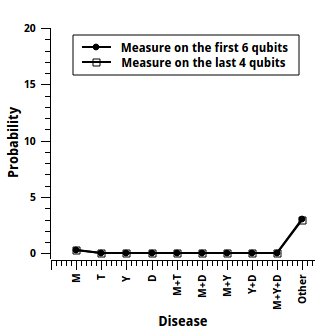}}
 \caption{Recognition efficiency of the diagnosis with only one particular 
symptom of malaria and the three common symptoms. M: malaria; T: typhoid fever; 
Y: yellow fever; D: dengue; Other: other disease. The graph shows the 
recognition efficiency when phase inversion is introduced (a) and when it is 
not 
introduced (b). The line with circles is the recognition efficiency when the 
measure is done on the first $6$ qubits (symptoms) while the line with empty 
square is the recognition efficiency when the measure is done on the last $4$ 
qubits (disease)).}
\label{fig:resul}
\end{center}
\end{figure}
\item Two particular symptoms of malaria, two particular symptoms of typhoid 
fever and one symptom that is only common to malaria and typhoid are introduced 
(Table \ref{tab:resul1} and Figure \ref{fig:resul1}).
\begin{table}[htpb]
  \begin{center}
  {%
\newcommand{\mc}[3]{\multicolumn{#1}{#2}{#3}}
\begin{center}
\begin{tabular}{llrl}\hline
Method & $\Lambda$ & \mc{2}{c}{$P_c/P_w$}\\\cline{3-4}
 &  & Measure on the $6$ first qubits & Measure on the $4$ last qubits\\\hline
with phase inversion & $17$ & $0.0650$ & $2.0089$\\
without phase inversion & $2372$ & $0.0348$ & $1.9408$\\\hline
  \end{tabular}
  \end{center}
}%
\caption{Relevant parameters of the \textbf{QAM-C2} method in case of the
diagnosis of malaria with typhoid fever.}
\label{tab:resul1}
\end{center}
 \end{table}
 \begin{figure}[hbt]
\begin{center}
\leavevmode
 \subfloat[with phase inversion]{
 \includegraphics[scale=0.8]{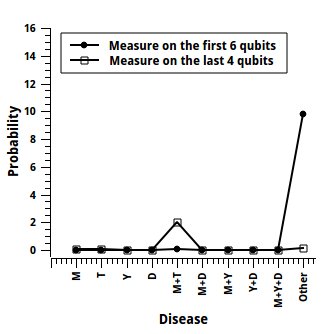}}\vspace{0.5cm}
 \subfloat[without phase inversion]{
 \includegraphics[scale=0.8]{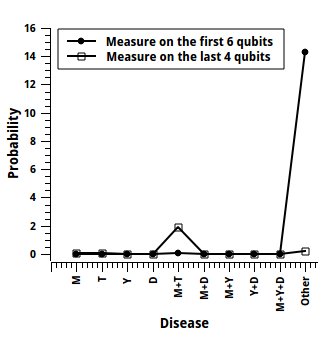}}
 \caption{Recognition efficiency of the diagnosis with two particulars symptoms
of malaria and typhoid and one particular common symptom of the two diseases. 
M: 
malaria; T: typhoid fever; Y: yellow fever; D: dengue; Other: other disease. 
The 
graph shows the recognition efficiency when phase inversion is introduced (a) 
and when it is not introduced (b). The line with circles is the recognition 
efficiency when the measure is done on the first $6$ qubits (symptoms) while 
the 
line with empty square is the recognition efficiency when the measure is done 
on 
the last $4$ qubits (disease)).}
\label{fig:resul1}
\end{center}
\end{figure}
\end{enumerate}

s

\subsection*{Overview on the simulation}
\begin{figure}[htpb]
\begin{center}
 \includegraphics[scale=0.45]{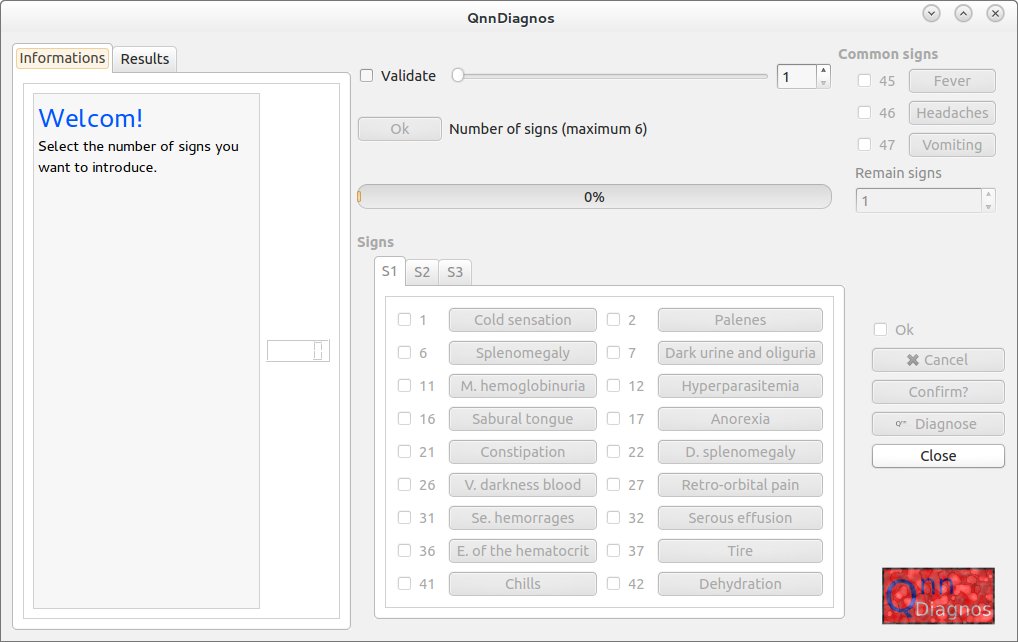}
 \caption{The interface of the software has a tab which gives information on 
what happens or on what should be done; a tab which gives the results of the 
diagnosis; and finally a set of buttons to introduce data and run the 
simulation 
to have the diagnosis.}
 \label{fig:qnndiagnos}
\end{center}
\end{figure}

All the simulation and results were made by writing the algorithms in 
\verb|C++| 
 language. The input register is the first $6$ qubits which compute symptoms,  
while the output register is the last $4$ qubits (see the database given in 
Section \ref{sec:signs}). A prototype of software, called \emph{QnnDiagnos} 
(Quantum neural networks for the Diagnosis), was designed in order to provide a 
user-friendly interface to physicians. It is developed with the open source 
version of \verb|C++| library \verb|Qt4| (see Figure \ref{fig:qnndiagnos}).

To use the software the physician first chooses the number of symptoms that he 
will introduce in the QAM (at least one symptom and a maximum of six symptoms), 
but it is not needful to introduce this number of symptom. This can be done 
before or after observations or discussion with the patient. Secondly, he 
introduces these symptoms, according to what he observes and the answers give 
by 
the patient, by clicking on the symptoms corresponding buttons on the 
interface. 
There are three tabs for symptoms of the four diseases ($S1$, $S2$ and $S3$). 
Thirdly, he obtains the diagnosis by comparing the recognition efficiency of 
the 
QAM for each disease gives in tab ''Results''. It is noteworthy that 
simulations 
are made on ''classical computer'', so each step of computation can be examined 
and recognition efficiencies calculated. So the disease with the greatest 
recognition efficiency can be view as the corresponding disease of the patient.

\section{Conclusion}\label{sec:concl}

\begin{figure}[hbtp]
\begin{center}
 \includegraphics[scale=0.9]{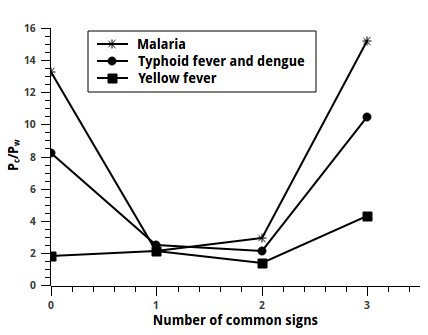}
 \caption{Evolution of recognition efficiency when one particular symptoms is
inserted with common symptoms for each disease.}
 \label{fig:accuracy}
\end{center}
\end{figure}

The use of quantum associative memory can be helpful to diagnose tropical 
diseases. The memory can distinguish single infection to multi-infection and do 
not need a lot of data to make the diagnosis; it needs few symptoms of disease 
and the common symptoms as shown on Figure \ref{fig:accuracy}. As shown in the 
recognition efficiency, the phase-inversion introduced in the original 
algorithm 
increases the capacity of the memory to make a good diagnosis. The memory can 
be 
a good alternative to help physicians without experience or laboratory to 
diagnose malaria, typhoid fever, yellow fever and dengue which are four 
tropical 
diseases sometime confused, using only clinical symptoms (some symptoms are 
clinical symptoms while others are biological symptoms which need 
laboratories). 
For future works, we plan to improve our database and increase the number of 
tropical diseases that the memory can recognize.

\appendix

\section{Symptoms of each group of diseases}\label{sec:signs}

 \begin{enumerate}
\subsection{Malaria}

\begin{multicols}{2}
 \item Cold sensation;
 \item Paleness;
 \item Severe anemia ($Hb<5\opn{g}/\opn{dl}$ or $Ht<15\%$);
 \item Profuse sweat;
 \item Stiffeness;
 \item Splenomegaly;
 \item Dark urine and oliguria (urine output$<400\opn{ml}$);
 \item Nausea;
 \item Pulmonary edema;
 \item Spontaneous hemorrhage;
 \item Macroscopic hemoglobinuria;
 \item Hyperparasitemia (parasitic density $>5\%$);
 \item Repeated generalize convulsions.
\end{multicols}

\subsection{Typhoid fever}
\begin{multicols}{2}
  \item Relative bradycardia;
  \item Epistaxis;
  \item Saburral tongue;
  \item Anorexia;
  \item Typhoid state;
  \item Rumble in the right iliac cavity;
  \item Yellowy diarrhea;
  \item Constipation;
  \item Delicate splenomegaly;
  \item Abdominal pinkish marks.

\end{multicols}

\subsection{Yellow fever}
\begin{multicols}{2}
  \item Iterus (jaundice);
  \item Renal troubles;
  \item Vomiting of darkness blood.

\end{multicols}

\subsection{Dengue}
\begin{multicols}{2}
 \item Retro-orbital pain;
 \item Leukopenia;
 \item Hemorrhagic manifestations;
 \item Positive Rumple-Leede phenomenon;
 \item Severe hemorrhages;
 \item Serous effusion;
 \item Maculopapular eruption;
 \item Articular pains;
 \item Thrombopenia (platelets $<100000/\opn{mm}^3$);
 \item Elevation of the hematocrit $\geq 20\%$ of is normal value.
\end{multicols}

\subsection{Common to the malaria and the typhoid fever}
\begin{multicols}{2}
 \item Tire;
 \item Abdominal pains;
 \item Hepatomegaly.
\end{multicols}

\subsection{Common to the malaria and the dengue}
\item Shock.

\subsection{Common to the malaria and the yellow fever}
\begin{multicols}{2}
 \item Chills;
 \item Dehydration.
\end{multicols}

\subsection{Common to the yellow fever and the dengue}
\item Muscular pains.

\subsection{Common to the malaria, the yellow fever and the dengue}
\item Trouble of the consciousness.

\subsection{Others (common to the four diseases)}
\begin{multicols}{2}
 \item Fever ($>38$\textcelsius);
 \item Headaches;
 \item Vomiting.
\end{multicols}

 \end{enumerate}

\end{document}